%% file: MainPaper.tex
\documentclass[letterpaper, 10 pt, conference]{ieeeconf}  

\IEEEoverridecommandlockouts                              

\usepackage{booktabs} 
\usepackage{url}
\usepackage{subfig}
\usepackage{graphicx}
\graphicspath{{figures/}}
\usepackage[english]{babel}
\usepackage[autostyle=true,english=american]{csquotes}
\usepackage{gensymb}
\usepackage{supertabular}
\usepackage[hidelinks]{hyperref}
\usepackage{amsmath}

\usepackage{acronym}
\newacro{AI}[AI]{Artificial Intelligence}
\newacro{UI}[UI]{user interface}
\newacro{GUI}[GUI]{graphical user interface}
\newacro{TLX}[TLX]{NASA-Task Load Index}
\newacro{RTLX}[Raw-TLX]{NASA Raw-Task Load Index}
\newacro{ER}[ER]{error rate}
\newacro{TCT}[TCT]{task completion time}
\newacro{HCI}[HCI]{Human-Computer Interaction}
\newacro{UX}[UX]{user experience}
\newacro{HFE}[HFE]{Human Factors and Ergonomics}
\newacro{cuDNN}[cuDNN]{CUDA Deep Neural Network library}
\newacro{RMSE}[RMSE]{root mean squared error}
\newacro{HMD}[HMD]{Head-Mounted Display}
\newacro{RF}[RF]{Random Forest}
\newacro{GP}[GP]{Gaussian process, long-plural = Gaussian processes}
\newacro{KNN}[\textit{k}NN]{\textit{k}-nearest neighbor}
\newacro{NN}[NN]{Neural Network}
\newacro{DNN}[DNN]{Deep Neural Network}
\newacro{CNN}[CNN]{Convolutional Neural Network}
\newacroplural{CNN}[CNNs]{Convolutional Neural Networks} 
\newacro{FCL}[FCL]{fully connected layer}
\newacro{BoD}[BoD]{Back-of-Device}
\newacro{FOV}[FoV]{field of view}
\newacro{RW}[RW]{Real World}
\newacro{IFRC}[IFRC]{index finger ray cast}
\newacro{FRC}[FRC]{forearm ray cast}
\newacro{EFRC}[EFRC]{eye-finger ray cast}
\newacro{HRC}[HRC]{Human-Robot Collaboration}
\newacro{HRI}[HRI]{Human-Robot Interaction}
\newacro{6DOF}[6DOF]{six-degree-of-freedom}
\newacro{LOOCV}[LOOCV]{leave-one-out cross-validation}
\newacro{CV}[CV]{cross-validation}
\newacro{RM}[RM]{repeated measure}
\newacro{ANOVA}[ANOVA]{Analysis of Variance}
\newacro{RMANOVA}[RM-ANOVA]{rRepeated Measures Analysis of Variance}
\newacro{AGATe}[AGATe]{AGreement Analysis Toolkit}
\newacro{GHoST}[GHoST]{Gesture Heatmap Toolkit Gesture Heatmaps Toolkit}
\newacro{GREAT}[GREAT]{Gesture Relative Accuracy Toolkit}
\newacro{GRT}[GRT]{Gesture Recognition Toolkit}
\newacro{DTW}[DTW]{Dynamic Time Warping}
\newacro{LHRD}[LHRD]{large high resolution display}
\newacro{GEQ}[GEQ]{Game Experience Questionnaire}
\newacro{SPGQ}[SPGQ]{Social Presence Gaming Questionnaire}
\newacro{JND}[JND]{just-noticeable difference}
\newacro{SUS}[SUS]{system usability scale}
\newacro{CSCW}[CSCW]{computer-supported cooperative work}
\newacro{CAD}[CAD]{computer-aided design}
\newacro{MR}[MR]{Mixed Reality}
\newacro{CVE}[CVE]{Collaborative Virtual Environment}
\newacro{AR}[AR]{Augmented Reality}
\newacro{AV}[AV]{Augmented Virtuality}
\newacro{VR}[VR]{Virtual Reality}
\newacro{PRISMA}[PRISMA]{Preferred Reporting Items for Systematic Reviews}
\newacro{PRISMA-Scope}[PRISMA-ScR]{Meta-Analyses Extension for Scoping Reviews}
\newacro{TF-IDF}[TF-IDF]{Term Frequency-Inverse Document Frequency}
\newacro{TF}[TF]{Term Frequency}
\newacro{AVs}[AVs]{Automated Vehicles}
\newacro{eHMIs}[eHMIs]{external Human-machine interfaces}
\newacro{SAR}[SAR]{Spatial Augmented Reality}
\newacro{MAR}[MAR]{Mobile Augmented Reality}
\newacro{IFR}[IFR]{International Federation of Robotics}
\newacro{ADL}[ADL]{Activity of Daily Living}
\newacroplural{ADL}{Activities of Daily Living} 
\newacro{LED}[LED]{Light-Emitting Diode}
\newacro{DoF}[DoF]{Degree-of-Freedom}
\newacroplural{DoF}[DoFs]{Degrees-of-Freedom}  
\newacro{HHC}[HHC]{Human-Human Collaboration}
\newacro{IDF}[IDF]{Inverse Document Frequency}
\newacro{QUEAD}[QUEAD]{Questionnaire for the Evaluation of Physical Assistive Devices}
\newacro{TiA}[TiA]{Trust in Automation Questionnaire}
\newacro{TOR}[TOR]{Take-Over-Request}
\newacro{ADMC}[ADMC]{Adaptive DoF Mapping Control}
\newacro{ML}[ML]{Machine Learning}
\newacro{IR}[IR]{infrared}
\newacro{IK}[IK]{inverse kinematics}
\newacro{ROS}[ROS]{Robot Operating System}
\newacro{TCP}[TCP]{Tool Center Point}
\newacro{DnD}[D\&D]{Design and Development}
\newacro{XR}[XR]{Extended Reality}
\newacro{CSV}[CSV]{Comma-separated values}
\newacro{PCA}[PCA]{Principal Component Analysis}
\newacro{RGBD}[RGB-D]{Red-Green-Blue-Depth}
\newacro{WHO}{World Health Organization}

\title{\LARGE \bf 
Adaptive Control in Assistive Application - A Study Evaluating Shared Control by Users with Limited Upper Limb Mobility
}

\ifx\draft\undefined
     \author{
        Felix Ferdinand Goldau$^{1,\circ}$ and Max Pascher$^{2,3,\circ}$ and Annalies Baumeister$^{4,\circ}$  and \\Patrizia Tolle$^4$ and Jens Gerken$^{2}$ and Udo Frese$^{1}$
        \thanks{$^{\circ}$ These authors contributed equally to this work.}
        \thanks{* This research is supported by the \textit{German Federal Ministry of Education and Research} (BMBF, FKZ: 
        \href{https://foerderportal.bund.de/foekat/jsp/SucheAction.do?actionMode=view&fkz=16SV8563}{16SV8563}, 
        \href{https://foerderportal.bund.de/foekat/jsp/SucheAction.do?actionMode=view&fkz=16SV8564}{16SV8564}, 
        \href{https://foerderportal.bund.de/foekat/jsp/SucheAction.do?actionMode=view&fkz=16SV8565}{16SV8565}, and 
        \href{https://foerderportal.bund.de/foekat/jsp/SucheAction.do?actionMode=view&fkz=01IW24001}{01IW24001}).}
        \thanks{$^{1}$Felix Ferdinand Goldau and Udo Frese are with the German Research Center for Artificial Intelligence (DFKI), 28359 Bremen, Germany
        {\tt\small felix.goldau@dfki.de, udo.frese@dfki.de}}%
        \thanks{$^{2}$Max Pascher and Jens Gerken are with the TU Dortmund University, Inclusive Human-Robot Interaction, 44227 Dortmund, Germany
        {\tt\small max.pascher@udo.edu, jens.gerken@udo.edu}}%
        \thanks{$^{3}$Max Pascher is also with the University of Duisburg-Essen, Human-Computer Interaction, 45127 Essen, Germany
        {\tt\small max.pascher@uni-due.de}}%
        \thanks{$^{4}$Annalies Baumeister and Patrizia Tolle are with the Frankfurt University of Applied Sciences, 60318 Frankfurt am Main, Germany
        {\tt\small annalies.baumeister@fb4.fra-uas.de, tolle@fb4.fra-uas.de}}%
    }
\else 
    \author{Blind for Review$^{1}$
    \thanks{*This work was not supported by any organization}
    \thanks{$^{1}$Blind for Review are with the University, City, Country
    {\tt\small firstname.lastname@blind.edu}}}
\fi

\begin{document}

\maketitle
\thispagestyle{empty}
\pagestyle{empty}


\begin{abstract}
\input{content/00-abstract.tex}

\end{abstract}



\input{content/01-introduction}
\input{content/02-related-work}
\input{content/03-approach}
\input{content/04-study}
\input{content/05-results}
\input{content/06-discussion}

\input{content/07-summary-and-conclusion}

\ifx\draft\undefined
    \section*{ACKNOWLEDGMENT}
    \noindent Our study was approved by the Ethics Committee of the \textit{Faculty of Business Administration and Economics of the University of Duisburg-Essen}.
\fi
\bibliographystyle{IEEEtran}
\bibliography{MainPaper}

\end{document}

%% file: content/00-abstract.tex



Shared control in assistive robotics blends human autonomy with computer assistance, thus simplifying complex tasks for individuals with physical impairments. This study assesses an adaptive Degrees of Freedom control method specifically tailored for individuals with upper limb impairments. It employs a between-subjects analysis with 24 participants, conducting 81 trials across three distinct input devices in a realistic everyday-task setting. Given the diverse capabilities of the vulnerable target demographic and the known challenges in statistical comparisons due to individual differences, the study focuses primarily on subjective qualitative data. The results reveal consistently high success rates in trial completions, irrespective of the input device used. Participants appreciated their involvement in the research process, displayed a positive outlook, and quick adaptability to the control system. Notably, each participant effectively managed the given task within a short time frame.


%% file: content/01-introduction.tex
\section{Introduction}\label{sec:introduction}

In 2023, the \ac{WHO} estimated that approximately 15\% of the global population lives with some form of disability~\cite{who.2023}, many of whom experience substantial, often permanent, reductions in limb usage. The resulting decreased mobility can severely restrict the ability to perform \acp{ADL} without external assistance, necessitating the almost constant presence of caregivers~\cite{Martinsen.2008}. However, constant caregiver presence is generally not desirable. Research by Pascher et al.\ demonstrated that individuals with physical disabilities strongly wish for personal space and alone-time~\cite{Pascher.2021recommendations}, which might be facilitated through the use of dependable robotic assistance~\cite{Pascher.2021recommendations}. Similarly, a comprehensive review by Kyrarini et al.\ highlighted the beneficial effects of assistive robotic technologies --- known as \emph{cobots} --- in aiding individuals with mobility issues~\cite{Kyrarini.2021survey}. Consequently, the decreased reliance on caregiver assistance supports the regaining of independence and addresses expressed wishes for self-determination.


However, introducing robots capable of (semi-)\,independent actions presents new challenges, potentially adding stress for the end-users if not properly considered during the design phase~\cite{Pollak.2020}. Pollak et al.~\cite{Pollak.2020} noted reduced sense of control felt by users when operating a cobot in autonomous mode, while switching to manual mode allowed participants to regain a sense of control and significantly lower stress levels. These findings are supported by Kim et al., who reported significantly higher satisfaction in the group using manual cobot control~\cite{Kim.2012}. Unlike routine tasks in industrial settings, such as assembly jobs~\cite{Braganca.2019}, the assistive care environment demands flexibility as cobots are tasked with a variety of support functions~\cite{Fattal.2019}. Managing robots in these scenarios remains demanding and requires continuous user involvement for efficient and safe system operation. A central issue arises from the types of robots employed, as multiple \acp{DoF} either require complex multidimensional input devices or involve time-extensive mode switching (e.g.,~\cite{Maheu.2011,Prattico.2021}). The former option is often unmanageable for individuals with mobility impairments, while the latter leads to increased task completion times~\cite{Herlant.2016modeswitch}. Consequently, these prevailing control strategies do not suit the needs of the intended audience.


In addressing this, adaptive \ac{DoF} control merges semi-autonomous operations with manual flexibility, dynamically adjusting a robot's \acp{DoF} for simplified interactions based on the environment. Introduced by Goldau \& Frese, this strategy enhances support for \acp{ADL}, outperforming traditional controls by using a \ac{CNN} to select optimal \acp{DoF} from real-time environmental feeds~\cite{Goldau.2021petra}. Further research by Pascher et al.\ demonstrated a reduction in mode switching, indicating a notable improvement over standard controls~\cite{Pascher.2024adaptix,Pascher.2023inTimeAndSpace,Kronhardt.2022adaptOrPerish} and explored different input devices for this adaptive control~\cite{Pascher.2024inputdevices}. Goldau \& Frese also confirmed the adaptive approach's advantages through heuristic behavior studies in a laboratory setting~\cite{heuristicControl}. Nonetheless, the real-world applicability and impact of these advances, especially in user studies targeting specific groups, are yet to be fully examined.
Building on these insights, the present study assesses the acceptance of adaptive control among the actual target group --- people with limited upper limb functionality --- through three select input devices.

\textbf{Our contribution} is two-fold: 1) we present a user study with the target group conducted at an international trade fair for rehabilitation and care, evaluating a novel shared control approach, and 2) provide an in-depth analysis of control performance data (average task completion time and average number of control switches) and subjective feedback (perceived workload, technology acquisition, and acceptance), highlighting the concepts' adaptability to various devices.

%% file: content/02-related-work.tex
\section{Assistive Robotics in Domestic Care}\label{sec:related}

When designing assistive technologies for vulnerable user groups, such as people with disabilities, efficient human-robot collaboration becomes paramount. Assistive robotics have the potential to significantly enhance independence and improve care by assisting and supplementing caregivers, thereby enhancing the quality of life for those in need~\cite{Kim.2012, Bemelmans.2012,Canal.2016,Hashimoto.2013}. Research attention has increasingly focused on how assistive robotic systems can assist individuals with motor impairments. Notably, projects like \textit{Robots for Humanity} led by Chen et al.~\cite{Chen.2013} and seminal studies like by Fattal et al.~\cite{Fattal.2019} explored the feasibility and user acceptance of these technologies. While the overarching aim is to fully integrate individuals with severe motor impairments into professional and social contexts, current assistive technologies predominantly target the performance of \acp{ADL}~\cite{Petrich.2022ADL}. These activities range from basic tasks like eating and drinking to more complex ones, including grooming and leisure activities~\cite{Chung.2013}.


Continual research efforts are expanding the capabilities of cobots and enhancing task performance. For instance, Gallenberger et al. utilized camera systems and machine learning in an autonomous robotic feeding system~\cite{Gallenberger.2019}, while Canal et al. introduced a learning-by-demonstration framework for feeding tasks~\cite{Canal.2016}. Both methods demonstrate how robotic arms can execute (semi-)autonomous tasks with minimal user intervention, thus underscoring the potential benefits of assistive technology. Implementing safe, user-friendly robotic solutions can fundamentally improve the quality of life for individuals needing assistance while ensuring that the user retains control~\cite{Llontop.2020quality}. This increased independence is particularly vital for those with motor impairments, reflecting their desire for more privacy and prolonged alone time~\cite{Park.2020}.


Drolshagen et al. found that individuals with disabilities readily adapt to working alongside cobots, even in close quarters~\cite{drolshagen2021acceptance}. Moreover, people with motor impairments tend to positively receive robotic assistance, especially when their specific needs are considered during the design process~\cite{Federici.2018}, and when sufficient oversight ensures a sense of security~\cite{Boada.2021}. Thus, effective communication of the robot's motion intent emerges as a crucial factor in achieving high acceptance among end-users~\cite{Pascher.2023robotMotionIntent}. These findings align with Beaudoin et al.'s investigation into the long-term usage of the Kinova Jaco, a notable advancement in assistive technology~\cite{Beaudoin.2019}.


\subsection{Shared Robot Control Applications}
\label{sec:sharedControl}

The appropriate level of autonomy in assistive robots attracts attention in current research. Highly autonomous systems (e.g.,~\cite{tsui2011want}), which minimize user interaction to mere oversight, can induce stress~\cite{Pollak.2020} and feelings of distrust among users~\cite{zlotowski2017can}. Conversely, for users with certain degrees of impairment, only minor adjustments to the users' otherwise manual control input~\cite{Sijs.2007} can pose significant challenges~\cite{Chen.2013,Kallinen.2017}. Shared control provides a middle ground by integrating manual user operation through standard input devices with algorithmic software assistance to adjust the resulting motion~\cite{Pascher.2024adaptix}. This approach effectively mitigates concerns associated with purely autonomous systems and manual controls~\cite{Abbink.2018}. In shared control, there is a collaborative effort between the user and the robot, empowering individuals with motor impairments to actively participate in their care. By balancing autonomy and user involvement, shared control systems offer a more acceptable and comfortable experience for individuals relying on assistive technologies~\cite{Reddy.2018,Gopinath.2017,Udupa.2021}.


A distinct approach is the adaptive \ac{DoF} control system proposed by Goldau \& Frese~\cite{Goldau.2021petra}. This system isolates the most likely \acp{DoF} of a robotic arm based on the current situation and aligns them with a low-\ac{DoF} input device. Effectively, this improves the classic mode-selection process by replacing the selectable modes with situation-adaptive directions of movement, allowing the user to easily control the arm. The process involves attaching a camera to the robotic arm's gripper and utilizing a \ac{CNN} trained on \acp{ADL} performed by individuals without motor impairments~\cite{Goldau.2021petra}, akin of the learning-by-demonstration method used in autonomous robots~\cite{Canal.2016}. Furthermore, this \ac{CNN}-based approach offers extensibility as it can be trained to distinguish between different situations, enhancing its practicality for everyday use. In their proof-of-concept study, which involved a 2D simulation environment featuring a robotic gripper representation and a target object, Goldau \& Frese observed faster task execution with the proposed system than manual controls. However, users perceived the shared control approach as complex, expressing a preference for a more extensive training phase, even in this low-\ac{DoF} environment. Their findings underscore the need for more intuitive and responsive interaction feedback when controlling the robot.

Pascher et al.'s \emph{\ac{ADMC}} concept draws inspiration from Goldau \& Frese's approach but extends it to three dimensions~\cite{Pascher.2024adaptix}. This extension increases the potential \acp{DoF}, enabling a more precise realization of \acp{ADL}. 
In their case studies, they show the advantages of an adaptive against a non-adaptive control approach~\cite{Kronhardt.2022adaptOrPerish,Pascher.2023inTimeAndSpace} and explored different input devices for the \ac{ADMC} concept~\cite{Pascher.2024inputdevices}. Following the transition to 3D, Goldau \& Frese expanded on their previous control by presenting a functional 3D prototype~\cite{heuristicControl}. Here, instead of generating the \acp{DoF} with a \ac{CNN}, they switched to a heuristic behavior-based approach. 
Using non-disabled participants in a laboratory environment, they showed the general viability of their control method, as well as the users' preferences for their novel approach.

However, as the adaptive \ac{DoF} control is yet to be evaluated with the target group in a realistic real-world scenario, its general accessibility and user acceptance remains to be assessed. Due to the diverse limitations of the target demographic, this accessibility coincides with a generalizability to different input devices.





%% file: content/03-approach.tex
\section{Technical Concept}
\label{sec:concept}

In line with the adaptive control principles discussed in Section~\ref{sec:sharedControl}, our study implemented a behavior-based heuristic control with a focus on assessing its applicability across various input devices and the acceptance within the targeted user group. 
The shared control approach adopted here is based on behaviors comparable to~\cite{heuristicControl}, albeit with a modification that incorporates known initial object poses. This adaptation aims to mitigate detection errors within the complex and cluttered environments typically encountered in trade fairs.

Unlike prior studies evaluating the usability of the proposed adaptive control concept~\cite{Kronhardt.2022adaptOrPerish,Pascher.2023inTimeAndSpace}, our experimental setup differs by concurrently integrating multiple approach- and graspable objects, as opposed to a single defined target. Our implementation is designed to operate without a predetermined sequence of actions, allowing users flexibility in interaction. To facilitate a practical assessment, we modified the research-oriented \emph{AdaptiX}~\cite{Pascher.2024adaptix} framework into a concise standalone \ac{ROS}-based system without a \ac{MR}-middleware. Similar to~\cite{heuristicControl}, our system uses a smart glass as a visualization interface, which we further augmented with options to use buttons or a joystick as input devices.


%% file: content/04-study.tex
\section{Study Method and Materials}
To assess the effectiveness of our adaptive control strategy, we conducted a supervised evaluation with a cohort of 24 participants. Our approach focused on qualitative data to gain individual insights into the broader implications of this diverse and challenging-to-generalize user group. Additionally, we supplemented this qualitative analysis by quantitative data derived from execution measurements and a \ac{RTLX} questionnaire~\cite{Hart.2006}.   

To achieve high external validity, we exclusively recruited participants from the target group and conducted the study in a relatively realistic environment, opting for a trade fair instead of an artificial laboratory setting. Participants used our adaptive control system with a designated input device to perform a simple task with a robot arm, after which they provided feedback on their experiences. The experiment primarily aimed to gather qualitative insights from the target group regarding the adaptive control strategy, supplemented by subjective questionnaires and performance data measurements.


\subsection{Study Design}\label{sec:study-design}
The study employed a between-subject design due to considerations of participant vulnerability (e.g., differing levels of fatigue) and diverse capabilities (e.g., only head-control being an option for some participants). Consequently, we used the input device as an independent variable, segmented into two distinct conditions: (1) \emph{Head-Control} and (2) \emph{Joystick}. Additionally, participants from both groups were asked to sample a third condition: (3) \emph{Assistive Buttons}.

The evaluated input devices were selected to be sufficiently distinct from one-another to accommodate a wide range of users, with the \emph{Joystick} and \emph{Head-Control} requiring finger and head dexterity respectively, whereas the \emph{Assistive Buttons} could be placed to be used with any body part. 
However, matching devices to participants' capabilities resulted in imbalanced data ($n_{\text{Head-Control}}=16$, $n_{\text{Joystick}}=8$, $n_{\text{Assistive Buttons}}=16$).
It is important to note that user familiarity with the devices varies greatly, as joysticks and buttons are more common than head-based controls. 

To facilitate an in-depth analysis of immediate user perceptions, we recorded both audio and video during the study. Additionally, we evaluated the following dependent variables:


\begin{itemize}
        \item \textbf{Average Task Completion Time:} The time to approach an object, pick it up, and position it at a designated target area was recorded (in seconds) for each trial.
        \item \textbf{Average Number of \ac{UI} Switches:} Within each trial, we documented instances of \ac{UI} switching, i.e., selections within the \ac{UI} independent of robot action, activated through a head-motion or button-press on the control device.
        \item  \textbf{Average Number of Mode Switches:} We measured mode switching, i.e., successful \ac{UI} switches followed by a user input to move the robot along a new \ac{DoF}.
        \item \textbf{Perceived Workload:} Following the completion of each condition, we assessed the six dimensions of the \ac{RTLX} questionnaire~\cite{Hart.2006} to gauge perceived workload. 
        \item \textbf{Level of Autonomy:} Upon completing all trials, we asked participants to identify their preferred level of autonomy on a Likert-scale 1--10 (1: manual control, 10: full autonomy).
\end{itemize}

Following the practical part of the study, we engaged participants with several open-ended questions to explore their experiences, understanding of the control method, interpretation of directional cues, and any significant issues they encountered. 

To extract participants' perceptions regarding the different control methods, the study's video and audio recordings were analyzed independently by three researchers through open coding. 
The resulting open codes were organized into affinity diagrams and further structured into themes, as detailed in Section~\ref{sec:thematicContentAnalysis}.


\subsection{Hypotheses}
Overall, we expect the adaptive control method to be well perceived by the target group, as long as the controls prove to be functional with the chosen input device. To assess this, we defined three hypotheses: 

\begin{enumerate}
 \item[\textbf{H1:}] After a short training, our target group of wheelchair-users with limited upper limb mobility is able to repeatedly use an adaptive \ac{DoF} control for a grasp-and-retrieve task.
 \item[\textbf{H2:}] Adaptive \ac{DoF} control is perceived as promising and accessible by the target group to perform tasks of \acp{ADL}.
 \item[\textbf{H3:}] The concept of adaptive control generalizes to different input devices.
 \end{enumerate}

\subsection{Apparatus}\label{sec:apparatus}

\begin{figure}
    \centering
    \includegraphics[width=\columnwidth]{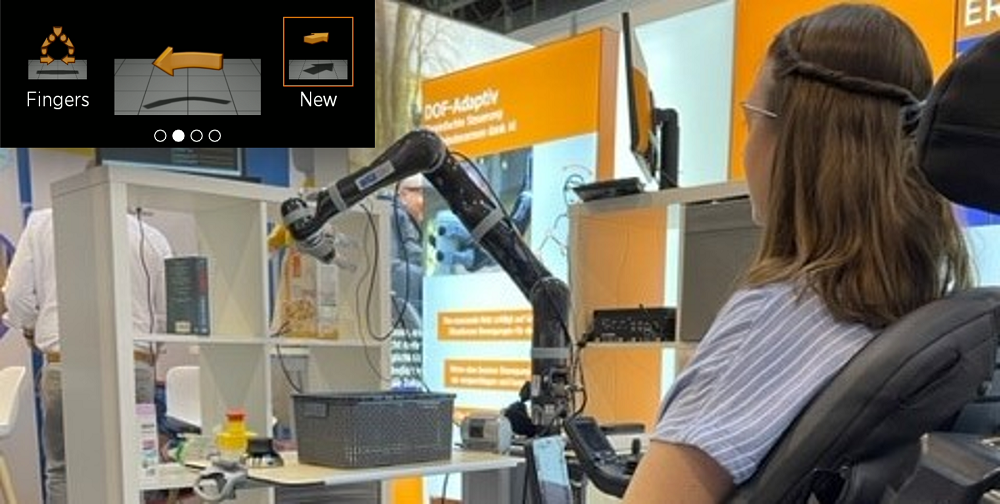}
    \caption{Study apparatus at the trade fair, illustrating the placement of user, table, and shelf, as well as the \ac{UI} visualized on the smart glasses (top left)}\label{fig:aufbau}
\end{figure}

\begin{figure}
    \centering
    \subfloat[]{\includegraphics[width=.32\columnwidth]{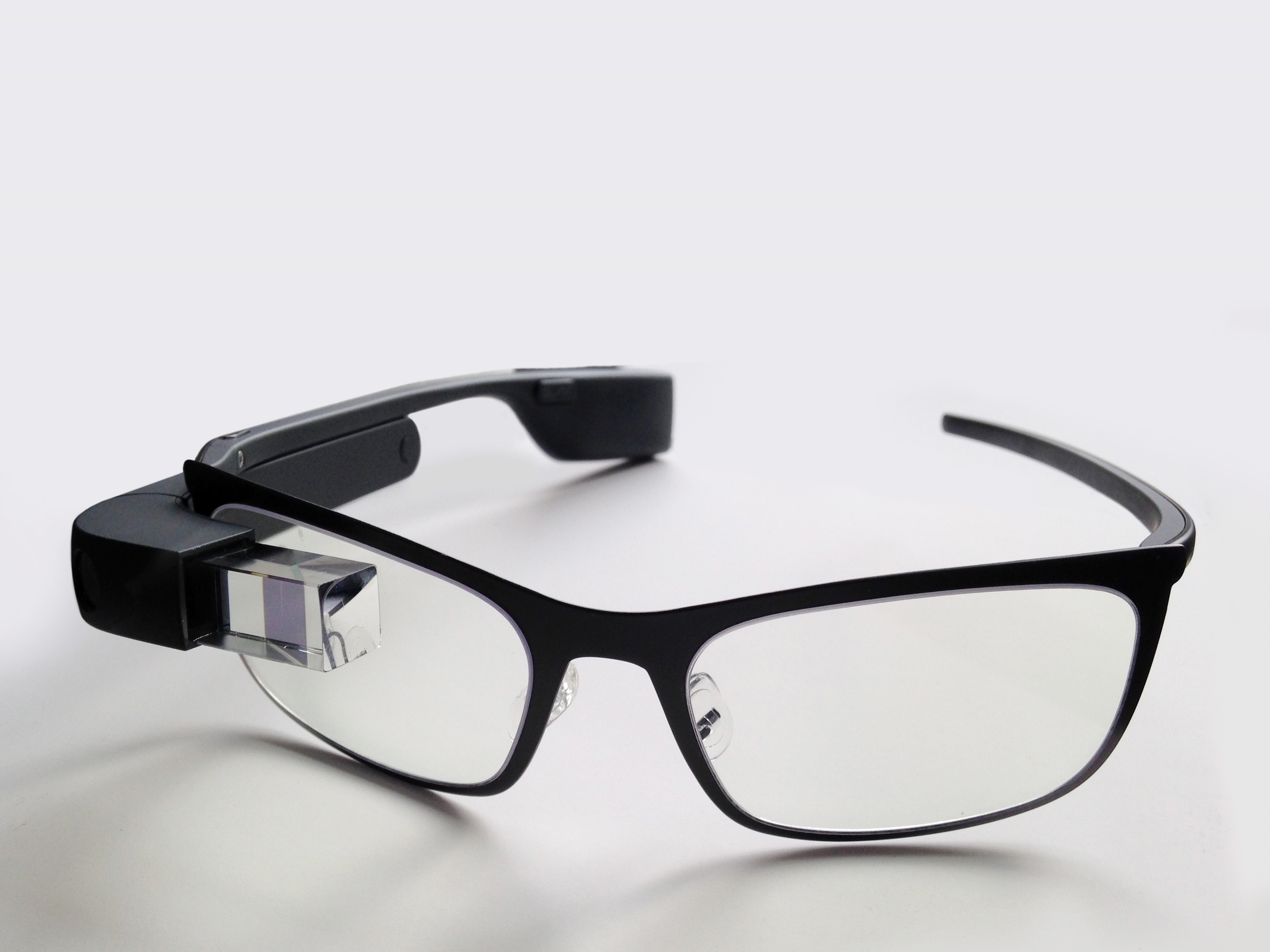}}
    \hfill
    \subfloat[]{\includegraphics[width=.32\columnwidth]{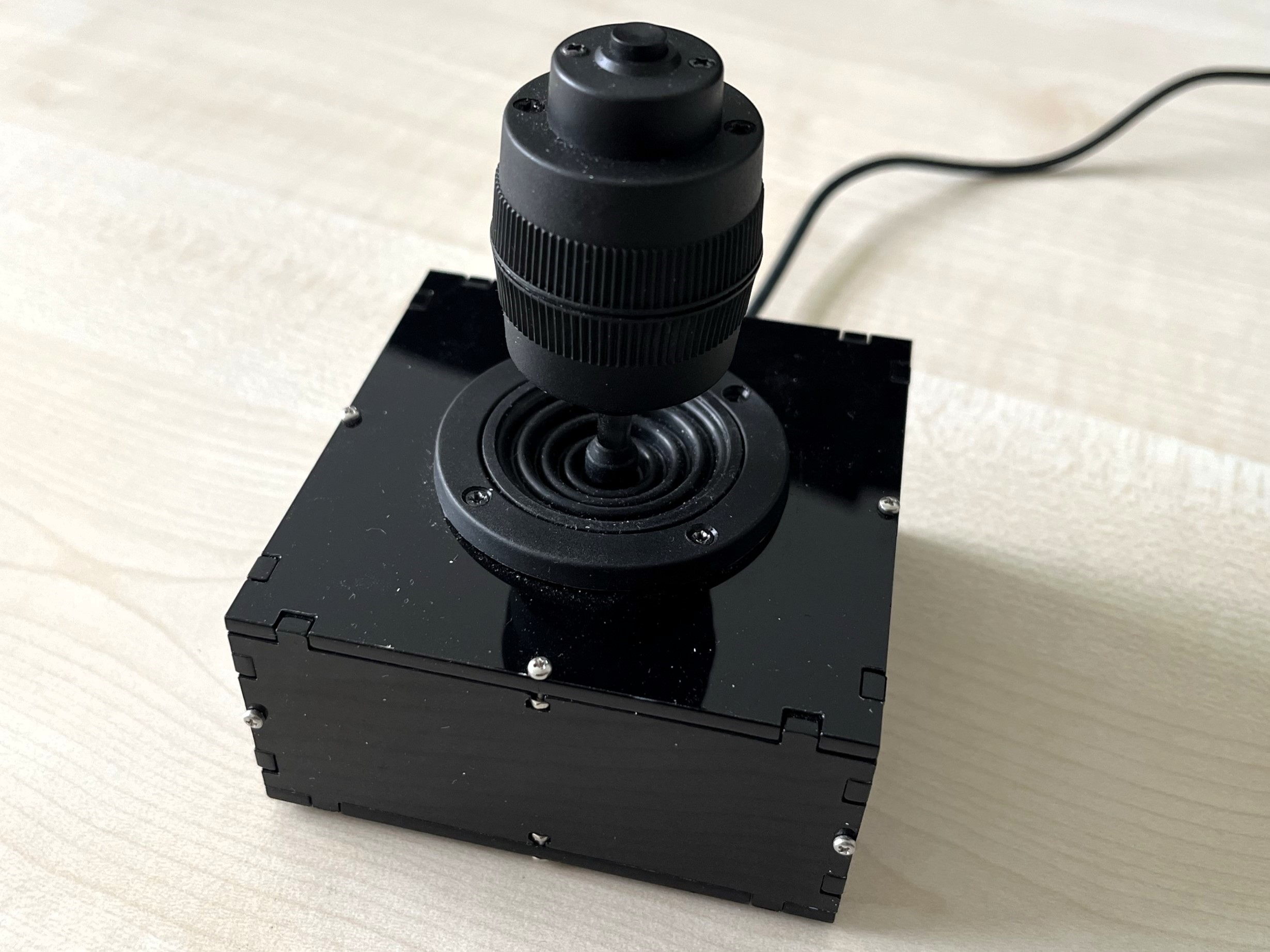}}
    \hfill
    \subfloat[]{\includegraphics[width=.32\columnwidth]{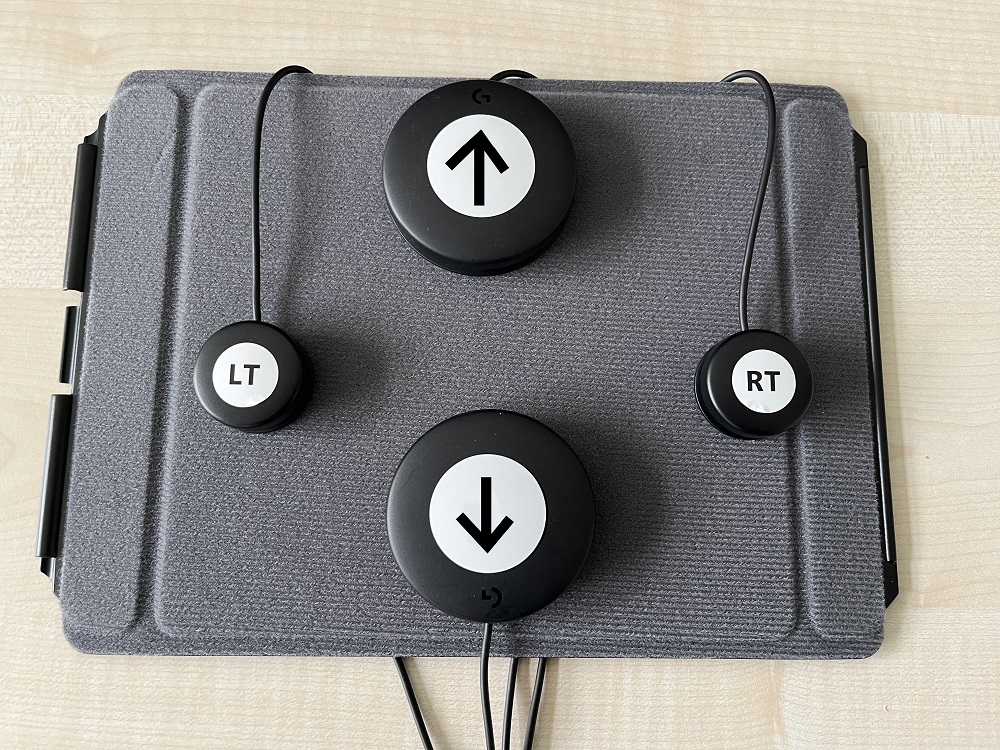}}
    
    \caption{Input devices used in the study: \textbf{(a)} \emph{Google Glass EE2}~\cite{glass_img}, \textbf{(b)} custom-built \emph{Joystick}, and \textbf{(c)} \emph{Assistive Buttons}}\label{fig:devices}
\end{figure}

The system was assembled on a mobile nightstand, simulating a setup typically found in nursing homes or hospitals (cf. Figure~\ref{fig:aufbau}). The central component was a \emph{Kinova Jaco Gen~2}~\cite{kinova} 7-\ac{DoF} assistive robotic arm with an \emph{Intel Realsense D435}~\cite{D435_Datasheet} color-and-depth-camera mounted to its end effector. 
As detailed in Section~\ref{sec:concept}, we evaluated multiple user input devices: A \emph{Google Glass EE2}~\cite{googleGlass} with a customized \emph{Munevo Drive}~\cite{munevo} software was used as the smart glasses, whereas an \emph{Xbox Adaptive Controller}~\cite{xboxcontroller} with external \emph{Assistive Buttons} and a custom-built \emph{Joystick} served as hand-controlled input devices. The devices are depicted in Figure~\ref{fig:devices}, with the \ac{UI}-visualization shown in the top left of Figure~\ref{fig:aufbau}.
To minimize external influences in the busy trade fair environment, all devices communicated via wired connections to a \ac{ROS}~\cite{ros} interface of an embedded \emph{Linux} computer. The only exception was the glasses, which were connected via a short-range \emph{Bluetooth} connection. 

\subsection{Participants}\label{sec:participants}
We focused on a target demographic of wheelchair users with reduced upper-limb mobility and the capability of wearing and using smart glasses. Consequently, individuals were excluded if they had vision impairments that made the glasses inaccessible or if the glasses did not fit (e.g., due to custom headrests). In total, 24 individuals --- 12 men and 12 women --- with varying motor impairments participated in the study. The age range of participants was 19 to 68 years, with an average age of 43.75 years (SD = 14.68). All participants relied on wheelchairs and had diverse health diagnoses, including spinal muscular atrophy, ALS, DMD, mitochondrial disease, AMC, MS, ICP, stroke, GNE myopathy, Charcot-Marie-Tooth syndrome, MMN, spina bifida, and generalized dystonia. One participant reported fully functional arms and hands, 20 had limited arm and hand function, and three had complete loss of arm and hand function.
Five participants had prior experience with an assistive robotic arm, another five had tested such an arm in the past, and 14 had no prior experience. Additionally, two individuals regularly used smart glasses, four had tried them before, and 18 had never used such glasses.

\subsection{Procedure}
\label{sec:procedure}

The study was conducted at the \emph{REHACARE},\footnote{\emph{REHACARE} trade fair.\ \url{https://www.rehacare.de}, last retrieved \today.} a leading international trade fair for rehabilitation and care in D{\"u}sseldorf, Germany. This location allowed for easy recruitment from the target group, as they are common visitors. The experiment setup was designed as part of a regular booth, with the participants facing a shelf to minimize visual distractions. 
Before the start, participants were thoroughly briefed about the research objectives and the to-be-completed tasks. Each participant provided explicit, informed consent to engage in the study and agreed to video/audio recording and documentation of all pertinent data.

The study administrator collected a socio-demographic questionnaire, monitored the experiment via a laptop, and provided instructions to participants on how to use the hardware and navigate the basic functions of the study environment. This followed the steps:

\begin{enumerate}
    \item The participant engages in a training trial with one-by-one assistance from the study administrator.
    \item 1--4 measurement trials (depending on individual capabilities) for the assigned condition are conducted.
    \item Based on personal capabilities, a subset of participants tested the \emph{Assistive Buttons} as an alternative input.
    \item Finally, we conducted a \ac{RTLX} questionnaire~\cite{Hart.2006} and a semi-structured interview.
\end{enumerate}

The study concluded with a debriefing session, with a total average session duration of 60 minutes. Participants were compensated with a 10 EUR food voucher for their time and engagement, a detail that was not disclosed beforehand.

\subsection{Experimental Setting and Task}\label{sec:task}


A small basket was placed as a target drop zone on a table in front the participant, thus allowing for a design that does not specify the object's final orientation. From the user's perspective, four objects were placed inside a 2x2 shelf behind the table. The robotic arm, attached to the table, could reach both the shelf spaces and the basket.
For each trial, the participants were tasked with guiding the robotic arm from its initial position to grasp an object from the shelf and put it into the basket. Upon successful placement, the trial concluded, and the object was removed from the basket. The robot, operated by the study administrator, returned to its starting point before starting a new trial for the remaining objects on the shelf. Neutral box-shaped objects were selected to prevent bias and ensure consistency across trials.


%% file: content/05-results.tex
\section{Results}\label{sec:results}
The study covers 81 measured trials (24 participants $\times$ 1--4 trials), with the training trials being excluded from analysis. To accommodate user capabilities, two thirds (16) of the participants evaluated the smart glass-based \emph{Head-Control}, while eight used the \emph{Joystick}. Additionally, 16 users evaluated \emph{Assistive Buttons} as a second input method after their first trails. 

\subsection{User Procedure Analysis}
As each trial begins at the same robot pose and involves only a single object with a pre-defined pose, we were able to analyze the user execution procedure for a singular task of approaching, grasping, and retrieving a single object. After excluding runs with external interruptions or major complications, we recorded 69~trials (36 \emph{Head-Control}, 17 \emph{Joystick}, 16 \emph{Assistive Buttons}).

\begin{figure}
    \centering
    \includegraphics[width=\columnwidth]{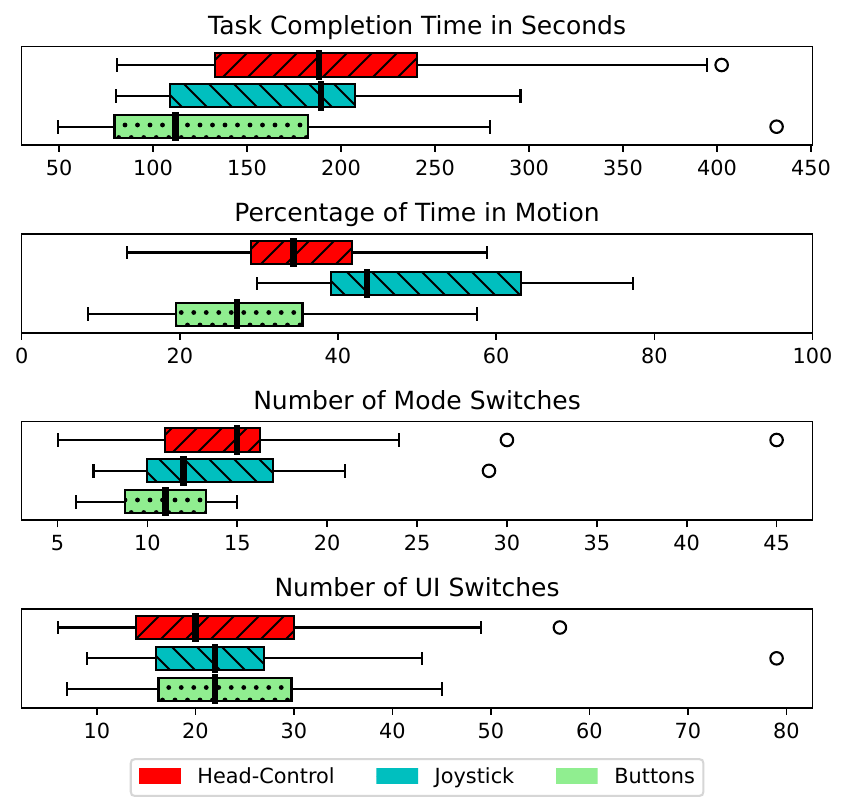}
    \caption{Box plot of execution procedure measurements over all users with $N_{\text{Head-Control}}=36$, $N_{\text{Joystick}}=17$, $N_{\text{Buttons}}=16$. The bold line represents the median}\label{fig:rosdata}
\end{figure}

\begin{figure}[hbt]
    \centering
    \includegraphics[width=\columnwidth]{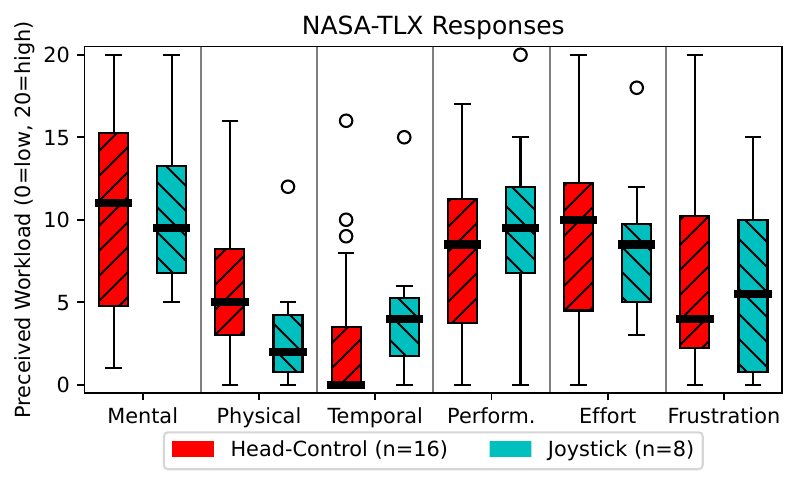}
    \caption{Results of the \acl{RTLX} questionnaires}\label{fig:nasaTLX}
\end{figure}

For each device, we recorded the overall task execution time, the percentage of time actually spend moving the robot, as well as the number of mode and \ac{UI} switches. 
Figure~\ref{fig:rosdata} shows an overview of the collected data for all users and devices. Figure~\ref{fig:nasaTLX} presents the subjective \ac{RTLX} scores. Each dimension is displayed as a box plot, separated by the type of control (\emph{Head-Control} or \emph{Joystick}) initially employed by the user. 

\subsection{Thematic Content Analysis}\label{sec:thematicContentAnalysis}
Throughout the trials and interviews, the participants verbalized their experiences, including challenges, moments of success, and improvement suggestions from their perspective. The audio recorded during these sessions was transcribed for thematic content analysis. Since participants were specifically asked to reflect on their study trials, the analysis primarily focuses on their individual experiences and perception of the adaptive control.

\subsubsection{Learning the Control}
Participants experienced varying learning curves, with only a minority finding the adaptive control intuitive at the beginning. Instead, most found the control to be initially strenuous, before improving their opinion after a short training period. Once \enquote{the concept was understood} (P11), usage became easier and more successful.


The primary challenge referenced by all participants was confronting the unfamiliar technology, especially using head movements in conjunction with smart glasses to control a robotic arm via augmented reality. Difficulties were noted even by those who used a joystick for input and the glasses solely for visualization. Participants had to quickly learn new skills and adjust to the adaptive control system, leading some to report feeling mentally overstrained at times.


The two participants already familiar with smart glasses and head movements to control their wheelchairs experienced the least difficulty in learning to control the robotic arm. In contrast, participants accustomed to using a robotic arm with a traditional joystick struggled to transfer their previous experience to the new system, regardless of whether they controlled the robot with head movements or a joystick during the trial. This difficulty was partly attributed to the nature of the shared control system, which imposed adaptive motions rather than the traditional fixed cardinal motions.


Given the recurring theme of technology acquisition and learning, participants were asked to estimate the training time required to use the robotic arm with adaptive control at home proficiently. While all participants were confident they could achieve proficiency with time and practice, their estimated training times varied, ranging from a few hours to several days or even weeks, potentially including further coaching sessions.


\subsubsection{Visualization}

A central feature of the control system is the arrow-based visualization displayed on the smart glasses (see top left in Figure~\ref{fig:aufbau}). Participants often felt uncertain about what action the robotic arm would take when following the arrow. As a result, they reported missing \enquote{the right one} (P8) and distrusting the system's suggestions. In these cases, frustration and uncertainty arose as the remaining task completion got more complicated and resulted in unexpected situations. 


Overall, participants generally understood simple arrow indications. However, more complex movements were challenging as many had difficulty with rotations, curved movements (multidimensional paths), and distinguishing between forward and backward control directions.
Some participants struggled to keep track of new suggestions and wanted the option for manual control in addition to the generated suggestions; an option that was not available during the study. Furthermore, participants also wished the system would indicate which object it was targeting.



\subsubsection{Physical Devices}
Participants using \emph{Head-Control} repeatedly forgot which physical movement corresponded to each \ac{UI} command, leading to periodic mix-ups. Conversely, those using the \emph{Joystick} struggled with the differences between controlling a robotic arm and a wheelchair. However, after training, the \emph{Joystick} users reported finding the adaptive control easy to use and an improvement over previously known controls.



Responses to the \emph{Head-Control} varied among participants. While most users found the adaptive control easier with increased insight and practice, two users experienced stress and physical tension from the head movements. One participant attributed this to their neuro-psychological impairment, finding the movements tiring and challenging to focus on. Nevertheless, participants generally found the adaptive control to be an interesting new method of controlling a robotic arm. Many described it as enjoyable once they became accustomed to it, with some even finding the suggestions and control to be intuitive once they \emph{got the hang of it} (P23).



Participants who also tested the \emph{Assistive Buttons} often found them to be the more accessible and more comfortable solution. Only one out of five participants from the \emph{Joystick} group who tested the \emph{Assistive Buttons} preferred the \emph{Joystick}. Among those who initially tested the \emph{Head-Control}, 12 tried the \emph{Assistive Buttons}, with only three preferring \emph{Head-Control}.
Participants who preferred the \emph{Assistive Buttons} found them more familiar and easier to use. They also found the limited direction options (left, right, forward, backward) to be more accessible.


\subsubsection{General User Remarks}

The study occurred at a fair rather than in a controlled laboratory environment, which was noted by participants as contributing to nervousness. Additionally, the bright light at the trade fair caused difficulties in recognizing graphics on the transparent display of the smart glasses. Participants found it strenuous to shift focus between the real robotic arm and the display and expressed a desire for visual alignment.
Despite these challenges, participants generally viewed the robotic system positively, appreciating the balance between suggestions and manual control.



\subsubsection{Preferred Level of Automation}

Given the frequent discussion surrounding the balance between automation and manual control in assistive robotics and shared control, participants were asked to express their own preferences regarding the level of automation on a scale ranging from 0 (i.e., no automation) to 10 (i.e., robotic system that functions completely autonomous).

\begin{figure}[hbt]
    \centering
    \includegraphics[width=\columnwidth]{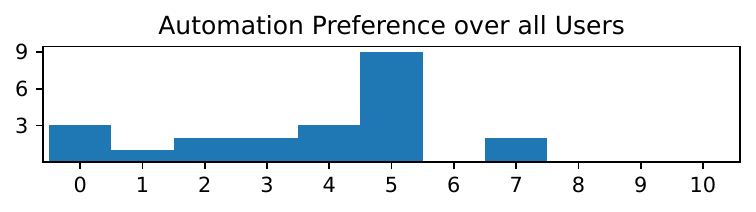}
    \caption{Histogram of preferred level of (robotic) automation of all users (0 complete manual control, 10 complete automation)}\label{fig:level_of_automation}
\end{figure}

The results, shown in Figure~\ref{fig:level_of_automation}, display a peak at the midpoint (5) with overall relatively evenly distributed responses at lower levels of automation. Notably, none of the participants favored the highest levels of automation. Their reluctance towards complete automation stemmed from the significance they placed on maintaining independence from technical devices. However, there was a consensus among users that some degree of automation would be beneficial or even essential, considering their physical limitations. Overall, users expressed a preference for support in manually controlling the robot rather than full automation.

%% file: content/06-discussion.tex
\section{Discussion}\label{sec:discussion}

Previous research~\cite{Pascher.2023inTimeAndSpace, Kronhardt.2022adaptOrPerish,Pascher.2024inputdevices} has demonstrated the general functionality of adaptive control with predetermined input devices, conducted by non-disabled users. The user procedure data generated in the present study corroborates these findings by confirming a general functionality independent of the input device. The measured completion times and number of mode or \ac{UI} switches showed no significant differences between input devices. Moreover, all participants completed the trials successfully and most evaluated the control concept positively. Consequently, these findings validate hypotheses H1 and H3 with a considerable degree of confidence.


In contrast, verifying H2 proves not as straightforward. Participants needed to learn the use of new technologies (glasses and adaptive control), which was mentally taxing and likely affected their perception of the adaptive control method. However, most participants did succeed after a brief period. Notably, they all anticipated that with more practice, usage would become easier, quicker, and more intuitive. This involves both learning the general concept of robot control and gaining a better understanding of the arrows and resulting robot motions. The latter, in particular, would lead to a clearer understanding of the robot's motion intent and encourage user trust in the suggestions and the overall system.


Overall, future users of an assistive system as used in this study could mitigate many --- if not all --- challenges experienced by the study participants. Among expected training effects, the glasses could be more personalized and calibrated more precisely to the individual user than is feasible in an experimental setup. Also, users are likely to select their preferred input device and become proficient with it. As shown by the preferred level of automation and user feedback, the presented concept fits requested the middle ground between automation and manual control. The results of the \ac{RTLX} indicate low physical and temporal demands when using adaptive \ac{DoF} control with an assistive robotic arm, thus representing an added value compared to previous solutions.





\subsection{Limitations}\label{sec:limitations}


This study evaluated a novel research-based shared control concept specifically with the intended user group, in contrast to much of the existing literature, which often includes participants outside this demographic. Our approach allows us to draw conclusions directly relevant to the end users without relying on generalizations from non-disabled user cohorts. However, to achieve this, the study took place at an international trade fair for rehabilitation and care, resulting in certain environment-specific limitations. 


Despite efforts to isolate trials and minimize external influences, the largely uncontrollable environment of the trade fair had a marked impact. However, even with the often audibly chaotic conditions, participants generally remaining focused during recordings, with only few getting noticeably distracted. Nevertheless, the noise and activity levels did affect the quality and options for recording quantitative data. As such, we focused more on the qualitative analysis of audible user comments during trials and their responses in the final interview. This approach yielded valuable insights, particularly because they came directly from the intended users themselves. Despite some distractions in the environment, they did not impact the qualitative data.

Finally, like most studies involving new control concepts, our participants only had a brief period to test the system. For comprehensive insights, the shared control approach requires extensive testing by target users in their everyday lives under assistive care settings.


%% file: content/07-summary-and-conclusion.tex
\section{Conclusions}\label{sec:conclusion}

This study presents an evaluation of a novel but literature-known concept of shared assistive robot control within the context of \acp{ADL} through direct engagement of the intended user demographic in a realistic setting. Our findings demonstrate the successful implementation of the control mechanism across multiple input devices, thereby highlighting its versatility and broad applicability. As such, the proposed control mechanism extends beyond a standalone solution and offers a significant enhancement over current best practices. 

Given that all study participants were representative of the target group, their quantitative feedback was particularly relevant and valuable. While some users initially encountered challenges with the system or found their assigned input device to be unfamiliar, all participants expressed confidence in being able to master the control with more time and practice. Notably, participants reported experiencing satisfaction in engaging with the presented control.

Overall, while our study does not conclusively show that adaptive control is straightforward to learn or intuitive, it does propose that the control method is indeed readily learnable within a short time frame, adaptable across different devices, and highly promising from an end-user perspective.



%% file: MainPaper.bbl
\begin{thebibliography}{10}
\providecommand{\url}[1]{#1}
\csname url@rmstyle\endcsname
\providecommand{\newblock}{\relax}
\providecommand{\bibinfo}[2]{#2}
\providecommand\BIBentrySTDinterwordspacing{\spaceskip=0pt\relax}
\providecommand\BIBentryALTinterwordstretchfactor{4}
\providecommand\BIBentryALTinterwordspacing{\spaceskip=\fontdimen2\font plus
\BIBentryALTinterwordstretchfactor\fontdimen3\font minus
  \fontdimen4\font\relax}
\providecommand\BIBforeignlanguage[2]{{%
\expandafter\ifx\csname l@#1\endcsname\relax
\typeout{** WARNING: IEEEtran.bst: No hyphenation pattern has been}%
\typeout{** loaded for the language `#1'. Using the pattern for}%
\typeout{** the default language instead.}%
\else
\language=\csname l@#1\endcsname
\fi
#2}}

\bibitem{who.2023}
\BIBentryALTinterwordspacing
{World Health Organization (WHO)}, ``Disability \& health report,'' 7 March
  2023. [Online]. Available:
  \url{https://www.who.int/news-room/fact-sheets/detail/disability-and-health}
\BIBentrySTDinterwordspacing

\bibitem{Martinsen.2008}
\BIBentryALTinterwordspacing
B.~Martinsen, I.~Harder, and F.~Biering-Sorensen, ``The meaning of assisted
  feeding for people living with spinal cord injury: a phenomenological
  study,'' \emph{Journal of Advanced Nursing}, vol.~62, no.~5, pp. 533--540,
  May 2008. [Online]. Available:
  \url{https://doi.org/10.1111/j.1365-2648.2008.04637.x}
\BIBentrySTDinterwordspacing

\bibitem{Pascher.2021recommendations}
\BIBentryALTinterwordspacing
M.~Pascher, A.~Baumeister, S.~Schneegass, B.~Klein, and J.~Gerken,
  ``Recommendations for the development of a robotic drinking and eating aid -
  an ethnographic study,'' in \emph{Human-Computer Interaction – INTERACT
  2021}, C.~Ardito, R.~Lanzilotti, A.~Malizia, H.~Petrie, A.~Piccinno,
  G.~Desolda, and K.~Inkpen, Eds.\hskip 1em plus 0.5em minus 0.4em\relax
  Springer, Cham, 2021. [Online]. Available:
  \url{https://doi.org/10.1007/978-3-030-85623-6\_21}
\BIBentrySTDinterwordspacing

\bibitem{Kyrarini.2021survey}
M.~Kyrarini, F.~Lygerakis, A.~Rajavenkatanarayanan, C.~Sevastopoulos, H.~R.
  Nambiappan, K.~K. Chaitanya, A.~R. Babu, J.~Mathew, and F.~Makedon, ``A
  survey of robots in healthcare,'' \emph{Technologies}, vol.~9, no.~1, p.~8,
  2021.

\bibitem{Pollak.2020}
\BIBentryALTinterwordspacing
A.~Pollak, M.~Paliga, M.~M. Pulopulos, B.~Kozusznik, and M.~W. Kozusznik,
  ``Stress in manual and autonomous modes of collaboration with a cobot,''
  \emph{Computers in Human Behavior}, vol. 112, p. 106469, 2020. [Online].
  Available:
  \url{https://www.sciencedirect.com/science/article/pii/S0747563220302211}
\BIBentrySTDinterwordspacing

\bibitem{Kim.2012}
D.-J. Kim, R.~Hazlett-Knudsen, H.~Culver-Godfrey, G.~Rucks, T.~Cunningham,
  D.~Portee, J.~Bricout, Z.~Wang, and A.~Behal, ``{How Autonomy Impacts
  Performance and Satisfaction: Results From a Study With Spinal Cord Injured
  Subjects Using an Assistive Robot},'' \emph{{IEEE Transactions on Systems,
  Man, and Cybernetics - Part A: Systems and Humans}}, vol.~42, no.~1, pp.
  2--14, 2012.

\bibitem{Braganca.2019}
\BIBentryALTinterwordspacing
S.~Bragan{\c{c}}a, E.~Costa, I.~Castellucci, and P.~M. Arezes, \emph{A Brief
  Overview of the Use of Collaborative Robots in Industry 4.0: Human Role and
  Safety}.\hskip 1em plus 0.5em minus 0.4em\relax Cham: Springer International
  Publishing, 2019, pp. 641--650. [Online]. Available:
  \url{https://doi.org/10.1007/978-3-030-14730-3\_68}
\BIBentrySTDinterwordspacing

\bibitem{Fattal.2019}
C.~Fattal, V.~Leynaert, I.~Laffont, A.~Baillet, M.~Enjalbert, and C.~Leroux,
  ``{SAM, an Assistive Robotic Device Dedicated to Helping Persons with
  Quadriplegia: Usability Study},'' \emph{{International Journal of Social
  Robotics}}, vol.~11, no.~1, pp. 89--103, 2019.

\bibitem{Maheu.2011}
V.~Maheu, P.~S. Archambault, J.~Frappier, and F.~Routhier, ``Evaluation of the
  jaco robotic arm: Clinico-economic study for powered wheelchair users with
  upper-extremity disabilities,'' in \emph{2011 IEEE International Conference
  on Rehabilitation Robotics}, 2011, pp. 1--5.

\bibitem{Prattico.2021}
\BIBentryALTinterwordspacing
F.~G. Pratticò and F.~Lamberti, ``Towards the adoption of virtual reality
  training systems for the self-tuition of industrial robot operators: A case
  study at kuka,'' \emph{Computers in Industry}, vol. 129, p. 103446, 2021.
  [Online]. Available:
  \url{https://www.sciencedirect.com/science/article/pii/S0166361521000531}
\BIBentrySTDinterwordspacing

\bibitem{Herlant.2016modeswitch}
L.~V. Herlant, R.~M. Holladay, and S.~S. Srinivasa, ``Assistive teleoperation
  of robot arms via automatic time-optimal mode switching,'' in \emph{The
  Eleventh ACM/IEEE International Conference on Human Robot Interaction}, ser.
  HRI '16.\hskip 1em plus 0.5em minus 0.4em\relax IEEE Press, 2016, p. 35–42.

\bibitem{Goldau.2021petra}
\BIBentryALTinterwordspacing
F.~F. Goldau and U.~Frese, ``Learning to map degrees of freedom for assistive
  user control: Towards an adaptive dof-mapping control for assistive robots,''
  in \emph{The 14th PErvasive Technologies Related to Assistive Environments
  Conference}, ser. PETRA 2021.\hskip 1em plus 0.5em minus 0.4em\relax New
  York, NY, USA: Association for Computing Machinery, 2021, p. 132–139.
  [Online]. Available: \url{https://doi.org/10.1145/3453892.3453895}
\BIBentrySTDinterwordspacing

\bibitem{Pascher.2024adaptix}
M.~Pascher, F.~F. Goldau, K.~Kronhardt, U.~Frese, and J.~Gerken, ``{AdaptiX –
  A Transitional XR Framework for Development and Evaluation of Shared Control
  Applications in Assistive Robotics},'' \emph{Proc. ACM Hum.-Comput.
  Interact.}, vol.~8, no. EICS, 2024, priprint on arXiv:
  \href{https://arxiv.org/abs/2310.15887}{https://arxiv.org/abs/2310.15887}.

\bibitem{Pascher.2023inTimeAndSpace}
M.~Pascher, K.~Kronhardt, F.~F. Goldau, U.~Frese, and J.~Gerken, ``{In Time and
  Space: Towards Usable Adaptive Control for Assistive Robotic Arms},'' in
  \emph{2023 32nd IEEE International Conference on Robot and Human Interactive
  Communication (RO-MAN)}.\hskip 1em plus 0.5em minus 0.4em\relax Institute of
  Electrical and Electronics Engineers (IEEE), 2023, pp. 2300--2307.

\bibitem{Kronhardt.2022adaptOrPerish}
K.~Kronhardt, S.~Rübner, M.~Pascher, F.~Goldau, U.~Frese, and J.~Gerken,
  ``Adapt or perish? exploring the effectiveness of adaptive dof control
  interaction methods for assistive robot arms,'' \emph{Technologies}, vol.~10,
  no.~1, 2022.

\bibitem{Pascher.2024inputdevices}
\BIBentryALTinterwordspacing
M.~Pascher, K.~Zinta, and J.~Gerken, ``{Exploring of Discrete and Continuous
  Input Control for AI-enhanced Assistive Robotic Arms},'' in \emph{Companion
  of the 2024 ACM/IEEE International Conference on Human-Robot Interaction},
  ser. HRI '24 Companion.\hskip 1em plus 0.5em minus 0.4em\relax New York, NY,
  USA: Association for Computing Machinery (ACM), 2024. [Online]. Available:
  \url{https://doi.org/10.1145/3610978.3640626}
\BIBentrySTDinterwordspacing

\bibitem{heuristicControl}
\BIBentryALTinterwordspacing
F.~Goldau and U.~Frese, ``Probabilistic combination of heuristic behaviors for
  shared assistive robot control,'' in \emph{The PErvasive Technologies Related
  to Assistive Environments (PETRA) conference (PETRA ’24)}.\hskip 1em plus
  0.5em minus 0.4em\relax New York, NY, USA: Association for Computing
  Machinery, 2024, p. 465–467. [Online]. Available:
  \url{https://doi.org/10.1145/3652037.3652071}
\BIBentrySTDinterwordspacing

\bibitem{Bemelmans.2012}
R.~Bemelmans, G.~J. Gelderblom, P.~Jonker, and L.~de~Witte, ``{Socially
  assistive robots in elderly care: a systematic review into effects and
  effectiveness},'' \emph{{Journal of the American Medical Directors
  Association}}, vol.~13, no.~2, pp. 114--120, 2012.

\bibitem{Canal.2016}
G.~Canal, G.~Aleny{\`a}, and C.~Torras, ``{Personalization Framework for
  Adaptive Robotic Feeding Assistance},'' in \emph{{Social Robotics}}.\hskip
  1em plus 0.5em minus 0.4em\relax Cham: {Springer International Publishing},
  2016, vol. 9979, pp. 22--31.

\bibitem{Hashimoto.2013}
K.~Hashimoto, F.~Saito, T.~Yamamoto, and K.~Ikeda, ``{A Field Study of the
  Human Support Robot in the Home Environment},'' in \emph{{2013 IEEE Workshop
  on Advanced Robotics and Its Social Impacts (ARSO)}}.\hskip 1em plus 0.5em
  minus 0.4em\relax Piscataway, NJ: IEEE, 2013, pp. 143--150.

\bibitem{Chen.2013}
T.~L. Chen, M.~Ciocarlie, S.~Cousins, P.~M. Grice, K.~Hawkins, K.~Hsiao, C.~C.
  Kemp, C.-H. King, D.~A. Lazewatsky, A.~E. Leeper, H.~Nguyen, A.~Paepcke,
  C.~Pantofaru, W.~D. Smart, and L.~Takayama, ``{Robots for humanity: using
  assistive robotics to empower people with disabilities},'' \emph{{IEEE
  Robotics {\&} Automation Magazine}}, vol.~20, no.~1, pp. 30--39, 2013.

\bibitem{Petrich.2022ADL}
L.~Petrich, J.~Jin, M.~Dehghan, and M.~Jagersand, ``A quantitative analysis of
  activities of daily living: Insights into improving functional independence
  with assistive robotics,'' in \emph{2022 International Conference on Robotics
  and Automation (ICRA)}, 2022, pp. 6999--7006.

\bibitem{Chung.2013}
C.-S. Chung, H.~Wang, and R.~A. Cooper, ``Functional assessment and performance
  evaluation for assistive robotic manipulators: Literature review,'' \emph{The
  Journal of Spinal Cord Medicine}, vol.~36, no.~4, pp. 273--289, 2013.

\bibitem{Gallenberger.2019}
D.~{Gallenberger}, T.~{Bhattacharjee}, Y.~{Kim}, and S.~S. {Srinivasa},
  ``Transfer depends on acquisition: Analyzing manipulation strategies for
  robotic feeding,'' in \emph{2019 14th ACM/IEEE International Conference on
  Human-Robot Interaction (HRI)}, 2019, pp. 267--276.

\bibitem{Llontop.2020quality}
D.~A. Rozas~Llontop, J.~Cornejo, R.~Palomares, and J.~A. Cornejo-Aguilar,
  ``Mechatronics design and simulation of anthropomorphic robotic arm mounted
  on wheelchair for supporting patients with spastic cerebral palsy,'' in
  \emph{2020 IEEE International Conference on Engineering Veracruz (ICEV)},
  2020, pp. 1--5.

\bibitem{Park.2020}
D.~Park, Y.~Hoshi, H.~P. Mahajan, H.~K. Kim, Z.~Erickson, W.~A. Rogers, and
  C.~C. Kemp, ``Active robot-assisted feeding with a general-purpose mobile
  manipulator: Design, evaluation, and lessons learned,'' \emph{Robotics and
  Autonomous Systems}, vol. 124, p. 103344, Feb 2020.

\bibitem{drolshagen2021acceptance}
S.~Drolshagen, M.~Pfingsthorn, P.~Gliesche, and A.~Hein, ``Acceptance of
  industrial collaborative robots by people with disabilities in sheltered
  workshops,'' \emph{Frontiers in Robotics and AI}, p. 173, 2021.

\bibitem{Federici.2018}
S.~Federici, \emph{Assistive technology assessment handbook}.\hskip 1em plus
  0.5em minus 0.4em\relax Boca Raton, FL: CRC Press, Taylor \& Francis Group,
  2018.

\bibitem{Boada.2021}
\BIBentryALTinterwordspacing
J.~P. Boada, B.~R. Maestre, and C.~T. Genís, ``The ethical issues of social
  assistive robotics: A critical literature review,'' \emph{Technology in
  Society}, vol.~67, p. 101726, 2021. [Online]. Available:
  \url{https://www.sciencedirect.com/science/article/pii/S0160791X21002013}
\BIBentrySTDinterwordspacing

\bibitem{Pascher.2023robotMotionIntent}
\BIBentryALTinterwordspacing
M.~Pascher, U.~Gruenefeld, S.~Schneegass, and J.~Gerken, ``How to communicate
  robot motion intent: A scoping review,'' in \emph{Proceedings of the 2023 CHI
  Conference on Human Factors in Computing Systems - CHI '23}, 2023. [Online].
  Available: \url{https://doi.org/10.1145/3544548.3580857}
\BIBentrySTDinterwordspacing

\bibitem{Beaudoin.2019}
M.~{Beaudoin}, J.~{Lettre}, F.~{Routhier}, P.~S. {Archambault}, M.~{Lemay}, and
  I.~{Gélinas}, ``{Long-term use of the JACO robotic arm: a case series},''
  \emph{Disability and Rehabilitation: Assistive Technology}, vol.~14, no.~3,
  pp. 267--275, 1 2019.

\bibitem{tsui2011want}
K.~M. Tsui, D.-J. Kim, A.~Behal, D.~Kontak, and H.~A. Yanco, ``“i want
  that”: Human-in-the-loop control of a wheelchair-mounted robotic arm,''
  \emph{Applied Bionics and Biomechanics}, vol.~8, no.~1, pp. 127--147, 2011.

\bibitem{zlotowski2017can}
J.~Z{\l}otowski, K.~Yogeeswaran, and C.~Bartneck, ``Can we control it?
  {A}utonomous robots threaten human identity, uniqueness, safety, and
  resources,'' \emph{International Journal of Human-Computer Studies}, vol.
  100, pp. 48--54, 2017.

\bibitem{Sijs.2007}
\BIBentryALTinterwordspacing
J.~Sijs, F.~Liefhebber, and G.~W.~R. Romer, ``Combined position \& force
  control for a robotic manipulator,'' in \emph{2007 {IEEE} 10th International
  Conference on Rehabilitation Robotics}.\hskip 1em plus 0.5em minus
  0.4em\relax {Institute of Electrical and Electronics Engineers (IEEE)}, 6
  2007. [Online]. Available: \url{https://doi.org/10.1109/icorr.2007.4428414}
\BIBentrySTDinterwordspacing

\bibitem{Kallinen.2017}
\BIBentryALTinterwordspacing
K.~Kallinen, ``The effects of transparency and task type on trust, stress,
  quality of work, and co-worker preference during human-autonomous system
  collaborative work,'' in \emph{Proceedings of the Companion of the 2017
  ACM/IEEE International Conference on Human-Robot Interaction}, ser. HRI
  '17.\hskip 1em plus 0.5em minus 0.4em\relax New York, NY, USA: Association
  for Computing Machinery (ACM), 2017, p. 153–154. [Online]. Available:
  \url{https://doi.org/10.1145/3029798.3038386}
\BIBentrySTDinterwordspacing

\bibitem{Abbink.2018}
\BIBentryALTinterwordspacing
D.~A. Abbink, T.~Carlson, M.~Mulder, J.~C.~F. de~Winter, F.~Aminravan, T.~L.
  Gibo, and E.~R. Boer, ``A topology of shared control
  systems{\textemdash}finding common ground in diversity,'' \emph{{IEEE}
  Transactions on Human-Machine Systems}, vol.~48, no.~5, pp. 509--525, Oct.
  2018. [Online]. Available: \url{https://doi.org/10.1109/thms.2018.2791570}
\BIBentrySTDinterwordspacing

\bibitem{Reddy.2018}
\BIBentryALTinterwordspacing
S.~Reddy, A.~Dragan, and S.~Levine, ``Shared autonomy via deep reinforcement
  learning,'' in \emph{Robotics: Science and Systems XIV}, ser. RSS2018.\hskip
  1em plus 0.5em minus 0.4em\relax Robotics: Science and Systems Foundation, 6
  2018. [Online]. Available: \url{http://dx.doi.org/10.15607/RSS.2018.XIV.005}
\BIBentrySTDinterwordspacing

\bibitem{Gopinath.2017}
D.~Gopinath, S.~Jain, and B.~D. Argall, ``Human-in-the-loop optimization of
  shared autonomy in assistive robotics,'' \emph{IEEE Robotics and Automation
  Letters}, vol.~2, no.~1, pp. 247--254, 2017.

\bibitem{Udupa.2021}
\BIBentryALTinterwordspacing
S.~Udupa, V.~R. Kamat, and C.~C. Menassa, ``Shared autonomy in assistive mobile
  robots: a review,'' \emph{Disability and Rehabilitation: Assistive
  Technology}, vol.~18, no.~6, p. 827–848, 6 2021. [Online]. Available:
  \url{http://dx.doi.org/10.1080/17483107.2021.1928778}
\BIBentrySTDinterwordspacing

\bibitem{Hart.2006}
S.~G. Hart, ``{Nasa-Task Load Index (NASA-TLX); 20 Years Later},''
  \emph{{Proceedings of the Human Factors and Ergonomics Society Annual
  Meeting}}, vol.~50, no.~9, pp. 904--908, 2006.

\bibitem{kinova}
\BIBentryALTinterwordspacing
{Kinova inc.}, ``Jaco assistive robot - user guide,'' Published online (last
  visited on \today), 2021, eN-UG-007-r05-202111. [Online]. Available:
  \url{https://assistive.kinovarobotics.com/uploads/EN-UG-007-Jaco-user-guide-R05.pdf}
\BIBentrySTDinterwordspacing

\bibitem{D435_Datasheet}
\BIBentryALTinterwordspacing
{Intel Corporation}, ``Intel realsense - product family d400 series,''
  Published online (last visited on 13.02.2023), 11 2022, revision 014.
  [Online]. Available:
  \url{https://www.intelrealsense.com/wp-content/uploads/2022/11/Intel-RealSense-D400-Series-Datasheet-November-2022.pdf}
\BIBentrySTDinterwordspacing

\bibitem{googleGlass}
Google, ``Glass enterprise edition 2,'' 2019.

\bibitem{munevo}
\BIBentryALTinterwordspacing
H.~Penkert, J.~C. Baron, K.~Madaus, W.~Huber, and A.~Berthele, ``Assessment of
  a novel, smartglass-based control device for electrically powered
  wheelchairs,'' \emph{Disability and Rehabilitation: Assistive Technology},
  vol.~16, no.~2, pp. 172--176, 2021. [Online]. Available:
  \url{https://doi.org/10.1080/17483107.2019.1646817}
\BIBentrySTDinterwordspacing

\bibitem{xboxcontroller}
Microsoft, ``Xbox adaptive controller,'' Online:
  \url{https://www.microsoft.com/en-us/d/xbox-adaptive-controller/8nsdbhz1n3d8}
  (last visited on \today), 2024.

\bibitem{ros}
\BIBentryALTinterwordspacing
{Stanford Artificial Intelligence Laboratory et al.}, ``Robotic operating
  system.'' [Online]. Available: \url{https://www.ros.org}
\BIBentrySTDinterwordspacing

\bibitem{glass_img}
\BIBentryALTinterwordspacing
{Mikepanhu}. (2014) Google glass with frame.jpg. {Wikimedia Commons}. Adapted
  under CC BY-SA 3.0 license. [Online]. Available:
  \url{https://upload.wikimedia.org/wikipedia/commons/b/be/Google_Glass_with_frame.jpg}
\BIBentrySTDinterwordspacing

\end{thebibliography}
